# Mathematical Model of Attraction and Repulsion Forces

Alexei Krouglov

*Matrox Graphics Inc.*

*3500 Steeles Ave. East, Suite 1300, Markham, Ontario L3R 2Z1, Canada*

Email: Alexei.Krouglov@matrox.com

This article represents the author’s personal view and not the view of Matrox Graphics Inc.

# ABSTRACT

Here I introduce the model in an attempt to describe the underlying reasons of attraction and repulsion forces between two physical bodies. Both electrical and gravitational forces are considered. Results are based on the technique developed in the Dual Time-Space Model of Wave Propagation.

## 1. Introduction

Developed recently [1] the Dual Time-Space Model of Wave Propagation (DTSMWP) proved to be a useful tool in investigating the wave nature of matter.

In present paper the model is applied to the phenomena of attraction and repulsion forces between physical bodies, which constitutes the subject of the field theory [2].

## 2. Model Assumptions

According to [1], the DTSMWP has the following assumptions.

In the time domain,

(1) The second derivative of energy's value with respect to time is inversely proportional to energy's disturbance.

(2) The first derivative of energy's level with respect to time is directly proportional to energy's disturbance.

In the space domain,

(3) The second derivative of energy's value with respect to direction is inversely proportional to energy's disturbance.

(4) The first derivative of energy's level with respect to direction is directly proportional to energy's disturbance.

## 3. Impact of Energy's Discrepancy in Space

I assume we have the jump of energy's value $U(x,t)$ at point $x_0$,

$$\begin{cases} U(x,t) = U_0 + \Delta E_0 & \textit{for } x \le x_0 \textit{ and } t \ge \tau > 0, \\ U(x,t) = U_0 & \textit{for } x > x_0 \textit{ and } t \ge \tau > 0. \end{cases} \quad (1)$$

I also assume that energy's level was initially stable,

$$\Phi(x,t) = U_0 \quad \textit{for } t < 0 \textit{ and } \forall x. \quad (2)$$

Then according to [1], we have the propagation of energy's disturbance in space.

Therefore we can write

$$\begin{cases} U(x_1,t) = U_0 \\ \Phi(x_1,t) = U_0 + \Delta E_0(x_1) \end{cases} \quad (3)$$

where $x_1 > x_0$, $t \ge \tau$, and $\Delta E_0(x_1)$ is the energy's disturbance propagated from point $x_0$ to point $x_1$.

From [1] we can see that energy's disturbance caused by the energy's level decreases in the space domain and retains the same sign.

Thus the following takes place,

$$\begin{aligned} &\Delta E_0 > \Delta E_0(x_1) > 0 \quad \textit{when } \Delta E_0 > 0, \\ &\Delta E_0 < \Delta E_0(x_1) < 0 \quad \textit{when } \Delta E_0 < 0. \end{aligned} \quad (4)$$

## 4. Compact Energy Body at Rest

Consider we have two jumps of the energy's values as follows,

$$\begin{cases} U(x,t) = U_0 & \textit{for } x < x_1 \textit{ and } t \geq \tau > 0, \\ U(x,t) = U_0 + \Delta E_1 & \textit{for } x_1 \leq x \leq x_2 \textit{ and } t \geq \tau > 0, \\ U(x,t) = U_0 & \textit{for } x > x_2 \textit{ and } t \geq \tau > 0. \end{cases} \tag{5}$$

and the energy's level was initially at rest,

$$\Phi(x,t) = U_0 \quad \textit{for } t < 0 \textit{ and } \forall x. \tag{6}$$

We assume that compact energy body lies at rest within $x_1 \leq x \leq x_2$.

We can conclude from [1] that there are forces $\overline{F}(x_1,t)$ and $\overline{F}(x_2,t)$ applied respectively to the points $x_1$ and $x_2$ of compact body with the following magnitudes,

$$\left|\overline{F}(x_1,t)\right| = \left|\overline{F}(x_2,t)\right| = \mu \cdot \left|\Delta E_1\right|, \tag{7}$$

where $t \geq \tau$ , and $\mu > 0$ is a constant.

To balance these forces we have to complement them with the opposite forces $\overline{F}'(x_1)$ and $\overline{F}'(x_2)$, therefore for $t > 0$,

$$\begin{cases} \overline{F}(x_1,t) + \overline{F}'(x_1) = 0, \\ \overline{F}(x_2,t) + \overline{F}'(x_2) = 0. \end{cases} \tag{8}$$

Note that forces $\overline{F}(x_1,t)$ and $\overline{F}(x_2,t)$ act in the direction of disturbance propagation, and forces $\overline{F}'(x_1)$ and $\overline{F}'(x_2)$ act in the opposite direction.

## 5. Impact of Energy's Discrepancy on Compact Energy Body

Let me consider three points of interest where we have the jumps of energy's values,

$$\begin{cases} U(x,t) = U_0 + \Delta E_0 & for\ x \le x_0\ and\ t \ge \tau > 0, \\ U(x,t) = U_0 & for\ x_0 < x < x_1\ and\ t \ge \tau > 0, \\ U(x,t) = U_0 + \Delta E_1 & for\ x_1 \le x \le x_2\ and\ t \ge \tau > 0, \\ U(x,t) = U_0 & for\ x > x_2\ and\ t \ge \tau > 0. \end{cases} \quad (9)$$

and the energy's level was initially at rest,

$$\Phi(x,t) = U_0 \quad for\ t < 0\ and\ \forall x. \quad (10)$$

We assume that the compact energy body lies at rest until the energy's disturbance from point $x_0$ reaches the point $x_1$ of compact body at time $t_1 > \tau$ .

Let me describe the forces applied to the point $x_1$ of compact body at the time $t = t_1 + \Delta t$ .

At first I will consider the case when magnitudes of energy's disturbances are close to each other, i.e. $|\Delta E_0| \approx |\Delta E_1|$, that causes inequality $|\Delta E_0(x_1)| < |\Delta E_1|$.

There are four possible situations.

(*a*) $\Delta E_0 > 0$ , $\Delta E_1 > 0$

$$\left|\overline{F}(x_1, t_1 + \Delta t)\right| = \mu \cdot (\Delta E_1 - \Delta E_0(x_1)) < \mu \cdot \Delta E_1 = \left|\overline{F}(x_1, t_1)\right| = \left|\overline{F}'(x_1)\right| . \quad (11)$$

Therefore the value of force $\overline{F}'(x_1)$, that has a direction inside the compact body, exceeds the value of an opposite force $\overline{F}(x_1, t_1 + \Delta t)$, and two bodies are repulsed.

(*b*) $\Delta E_0 < 0$, $\Delta E_1 > 0$

$$\left|\overline{F}(x_1, t_1 + \Delta t)\right| = \mu \cdot (\Delta E_1 - \Delta E_0(x_1)) > \mu \cdot \Delta E_1 = \left|\overline{F}(x_1, t_1)\right| = \left|\overline{F}'(x_1)\right|. \tag{12}$$

Therefore the bodies are attracted.

(*c*) $\Delta E_0 > 0$, $\Delta E_1 < 0$

$$\left|\overline{F}(x_1, t_1 + \Delta t)\right| = \mu \cdot \left|\Delta E_1 - \Delta E_0(x_1)\right| > \mu \cdot \left|\Delta E_1\right| = \left|\overline{F}(x_1, t_1)\right| = \left|\overline{F}'(x_1)\right|. \tag{13}$$

Therefore the bodies are attracted.

(*d*) $\Delta E_0 < 0$, $\Delta E_1 < 0$

$$\left|\overline{F}(x_1, t_1 + \Delta t)\right| = \mu \cdot \left|\Delta E_1 - \Delta E_0(x_1)\right| < \mu \cdot \left|\Delta E_1\right| = \left|\overline{F}(x_1, t_1)\right| = \left|\overline{F}'(x_1)\right|. \tag{14}$$

Therefore the bodies are repulsed.

Thus we have described so far the forces of electrical attraction and repulsion.

Now I will consider the case when $\Delta E_0 >> \Delta E_1 > 0$. From here we have an inequality $\Delta E_0(x_1) > 2 \cdot \Delta E_1 > 0$, that as we can see soon, creates the force of gravitational attraction,

$$\left|\overline{F}(x_1, t_1 + \Delta t)\right| = \mu \cdot \left|\Delta E_1 - \Delta E_0(x_1)\right| > \mu \cdot \Delta E_1 = \left|\overline{F}(x_1, t_1)\right| = \left|\overline{F}'(x_1)\right|. \tag{15}$$

Hence the bodies are attracted.